\documentclass[11pt]{article}
\usepackage{amsmath}
\usepackage{graphicx}
\usepackage{amsfonts}
\usepackage{amssymb}
\usepackage{epsf}
\usepackage{latexsym}

\def\be{\begin{equation}}
\def\ee{\end{equation}}
\def\bea{\begin{eqnarray}}
\def\eea{\end{eqnarray}}
\def\pa{\partial}

\def\fn{\footnote}
\def\case#1/#2{\textstyle\frac{#1}{#2}}

\begin{document}
\begin{titlepage}

\vspace{.7in}

\begin{center} 
\Large
{\bf THE CAMPBELL--MAGAARD THEOREM IS INADEQUATE} 
\normalsize

\vspace{0.1in}

\Large{\bf AND INAPPROPRIATE AS A PROTECTIVE THEOREM}
\normalsize

\vspace{0.1in}

\Large{\bf FOR RELATIVISTIC FIELD EQUATIONS} 
\normalsize

\normalsize 

\vspace{.2in}

{\bf Edward Anderson}$^*$

\vspace{.2in}

{\em  Astronomy Unit, School of Mathematical Sciences, 
            Queen Mary, University of London, UK}

\normalsize

\vspace{.2in}

\baselineskip=24pt

\vspace{.2in}

\begin{abstract}

Given a particular prescription for the Einstein field equations, it is important 
to have general protective theorems that lend support to it.  The prescription of 
data on a timelike hypersurface for the ($n$ + 1)-d Einstein field equations arises 
in `noncompact Kaluza--Klein theory', and in certain kinds of braneworlds and 
low-energy string theory.  The Campbell--Magaard theorem, which asserts local 
existence (and, with extra conditions, uniqueness) of analytic embeddings of completely general $n$-d manifolds 
into vacuum ($n$ + 1)-d manifolds, has often recently been invoked as a protective theorem 
for such prescriptions.  

But in this paper I argue that there are problems with loosening the Campbell--Magaard theorem of restricitive 
meanings in its statement, which is a worthwhile thing to do in pursuit of the proposed applications.  
While I remedy some problems by 
identifying the required topology, delineating what `local' can be taken to mean, and offering a new, 
more robust and covariant proof, other problems with the proposed applications remain insurmountable.  
The theorem lends only inadequate support, both because it offers no guarantee of continuous dependence on 
the data and because it disregards causality.  Furthermore, the theorem is only for the 
analytic functions which renders it inappropriate for the study of the relativistic field 
equations of modern physics.  

Unfortunately, there are as yet no known general theorems that offer adequate protection to the 
proposed applications' prescription.  I conclude by making some suggestions for more modest 
progress.  

\end{abstract}

\vspace{.2in}

\end{center}

\noindent$^*$ \em Next addresses: as of Oct 2004, 
Peterhouse College Cambridge UK and DAMTP Cambridge UK; 

\noindent additionally, as of Jan 2005, 
Department of Physics, P-412 Avadh Bhatia Physics Laboratory, 
University of Alberta, Edmonton, Alberta, Canada.  

\mbox{ }

\normalfont

\noindent Email addresses: eda@maths.qmul.ac.uk, ea212@cam.ac.uk

\vspace{5in}

\end{titlepage}

%----------------------------------------------------------------------------------------------
\section{Introduction}
%----------------------------------------------------------------------------------------------

The Campbell--Magaard (CM) theorem \cite{Camp26, Mag63} concerns the existence (and with extra conditions uniqueness) 
%uniqueness if Magaard procedure is followed, and per choice of extrinsic curvature components. 
of embeddings of $n$-d manifolds into vacuum ($n$ + 1)-d manifolds. This theorem, qualified as 
`old and little known', has recently been 
claimed to lend support \cite{RTZ, STMbook, Claimsup, testsandsupp, AL, DMR, FigRom} to 
`space-time-matter' or `noncompactified Kaluza--Klein' (NKK) theory 
\cite{STMbook, Wearly, W90, NKKitself}.   
This is a proposed 5-d toy model for the investigation of the consequences of higher-d physics.  
In this article, I question the suitability of the CM theorem for such an application and hence 
I question the foundations of NKK theory.  Reasons to consider NKK seriously enough to ask such questions 
include 1) that one may be able to untangle and then understand some of the many effects of higher-d physics 
by such models, and 2) that NKK theory is playing a role in experimental predictions and proposals \cite{testsforNKK, testsandsupp}; 
it even plays a role in a recent cover feature in the New Scientist \cite{NewScientist}.    

Further reasons to question the applicability of the CM theorem are that  
it has also appeared in connection with constructing higher-d GR solutions \cite{LRT, LRRT}, 
low-energy string theory solutions \cite{Lidsey} and lower-d gravity \cite{LRRT}.  
The CM theorem generalizes \cite{AL, DR, ADLR} to a wider range of situations 
in higher-d physics, where cosmological constants and scalar fields are commonplace.  
The CM theorem and its generalizations have also been suggested \cite{SW03, DRbranes} as an 
underlying theorem relevant for building certain types\fn{By which I mean the well-known \cite{BWlit, SMSMar} and 
thick brane models (see e.g \cite{thickbranes}).} of braneworlds (see also 
\cite{Bron}), and have also been studied in various other contexts, e.g \cite{Maia, Romlit, harm, Novello, DMR, FigRom}.

In this paper, serious difficulties with using the CM theorem for these applications are revealed and studied.  
It is in the interest of the proposed applications to have protective results in extended regions.\fn{I carefully delineate the differences 
between Campbell's conjecture, the CM theorem  and the sort of result 
that would be desirable for the proposed applications.} Moreover, the historical development 
of the GR Cauchy and initial-value problems (Sec 2--4)\fn{I consider especially Darmois' 
work \cite{Darmois23, Darmois27} and Four\`{e}s-Bruhat's \cite{B52, B56}, which are the 
foundations for the GR CP and IVP literature (see \cite{CB62, HE, FMrev, 
CBY80, Wald, RFrev, Rrev, Cook, BS, 50} for more recent reviews).} provides both guidance and 
limitations on this pursuit.  I argue (Secs 2--4) that the CM theorem can be seen as a 
collection of workings each of which is of a well-known type from the roughly contemporary 
(very early) GR CP and IVP literature.  This was not picked up by Magaard \cite{Mag63}, 
by the historians \cite{EK1, EK2, Goenner} or in the introduction of the theorem into the 
modern literature \cite{RTZ}.  I consider limitations on the extendibility of Magaard's proof 
of the CM theorem toward a proof that would be of greater value for the proposed applications 
in Secs 4--7.   I provide improvements on Magaard's proof  in Secs 5-6, which include 
identifying the required topology, spelling out what the word `local' can and cannot be taken 
to mean in the statement of the theorem, and working covariantly.   However, other problems 
with such a program remain insurmountable (Sec 7).  The theorem offers only inadequate support 
both because it can not encompass continuous dependence on the data and because it disregards 
causality.  Furthermore, the theorem is solely about the analytic functions which renders it 
inappropriate for the study of the relativistic field equations of modern physics.  
Thus there are strong reasons why the CM theorem should not be taken seriously as a protective theorem 
for general relativistic field equations.  

I conclude in Sec 8.1 with the case against the CM theorem, and in Sec 8.2 with an 
account of how in the case of its applications proposed in the literature 
there is currently no other satisfactory general protective theorem known.   
Thus these proposed applications remain unprotected.  In Sec 8.3 I 
discuss the position of NKK theory in this light.  In Sec 8.4 I point out that 
the construction of particular examples in various claimed applications do not 
actually make contact with the CM theorem.  The methods actually used in these  
examples are often rather among the cruder simplifications of the well-known 
Lichnerowicz--York conformal method for the IVP 
\cite{Lich44, Yorklit, CBY80, discoftricks}.  Finally in Sec 8.5 I discuss some 
more limited ways to approach the braneworld application while staying within 
bona fide mathematical physics.

%----------------------------------------------------------------------------------------------
\section{Campbell and his contemporaries}
%----------------------------------------------------------------------------------------------

Campbell, while primarily a pure mathematician, was well aware of GR \cite{Campother}.  
%which he is said to have picked up from Eddington \cite{EK1}.  
However, his 1926 book presents what is a {\it split} of the ($n$ + 1)-d Euclidean version 
of the vacuum Einstein field equations (EFE's) ${\cal R}_{AB} = 0$ with respect to a $n$-d hypersurface 
%non-null irrelevant: Euclidean DEF
$\mu = const$ without any comments about, or applications to, GR.\fn{While Campbell intended to 
end his book with GR comments \cite{Campunfin}, he died in 1924 and the book 
remained unfinished in this respect.} The split is\fn{I adopt a modern notation throughout 
this paper.  I use capitals for 
($n$ + 1)-d indices and lower-case letters for $n$-d indices.  
$h_{ab}$ is the $n$-d metric, 
$D_a$ is the $n$-d covariant derivative,  
$R_{ab}$ is the $n$-d Ricci tensor,
$g_{AB}$ is the ($n$ + 1)-d metric,
$\nabla_A$ is the ($n$ + 1)-d covariant derivative and 
${\cal R}_{AB}$ is the 
($n$ + 1)-d Ricci tensor. }
\be
0 = {\cal R}_{ab} = R_{ab} + \epsilon 
\left(
\Omega_{ab}\Omega - 2{\Omega_{a}}^c\Omega_{bc}
\right) 
- \frac{\epsilon}{\phi}  \frac{\partial \Omega_{ab}}{\partial \mu} \quad 
+ \frac{1}{\phi}D_{a}D_{b}\phi
%=-\kappa \left( \tilde{T}_{ik}-\frac{1}{n-1}g_{ik}\tilde{T}\right) MATTER TERM IN ADLR NOTATION , 
\label{ab}
\mbox{ } ,
\ee
\be
0 = {\cal R}_{a\mu} = \frac{\epsilon }{\phi}
\left( 
D^{b}\Omega_{ab} - D_{a}\Omega
\right) 
\mbox{ } , 
%=-\kappa \tilde{T}_{i}^{y}, MATTER TERM IN ADLR NOTATION
\label{amu}
\ee
\be
0 = {\cal R}_{\mu\mu} =  \frac{\pa \Omega}{\pa \mu} - \epsilon \phi D^2\phi - \Omega_{ab}\Omega^{ab} 
\mbox{ } .
\label{mumu}
\ee
$$
\mbox{Here, } \mbox{\hspace{0.5in}}
g_{AB} = 
\left(
\begin{array}{ll}
h_{ab}           &
0                \\
0                &
\epsilon\phi^2   \\ 
\end{array}
\right)
\mbox{\hspace{0.5in}}  , \mbox{\hspace{0.5in}} 
\epsilon = 1 
\mbox{\hspace{0.5in}}  \mbox{and} \mbox{\hspace{0.5in}}
\Omega_{ab} = -\frac{1}{2\phi}\frac{\pa h_{ab}}{\pa\mu} 
\mbox{ } . \mbox{\hspace{0.5in}}
$$
He eventually identified $n + 1$ of these equations, (\ref{amu}) and 
$\left[
(\ref{mumu}) - \frac{1}{2} \mbox{ trace}(\ref{ab})
\right]$ i.e
\be
R - \Omega_{ab}\Omega^{ab} + \Omega^2  = 0 \mbox{ } , 
\label{Gauss}
\ee 
as being `different' from the other equations (\ref{ab}).  
So there are {\it two systems to treat separately}.  
He then proposed, but by no means proved, the following conjecture about the above split.  

\mbox{ }

\noindent{\bf Campbell's conjecture} Any $n$-space is `surrounded' by a vacuum ($n$ + 1)-space.

\mbox{ }

\noindent What he offered as a piece of a proof for this is that the `different' equations 
(\ref{amu}, \ref{Gauss}) propagate away from the original $\mu = const$ hypersurface.  

\mbox{ }

It is however illuminating to consider Campbell's work within its proper historical context.  
He appears to have been unaware that his relativist contemporaries were in the middle 
of learning about the conceptualizations of such splits.  In any case, Campbell's book 
and death pre-date the consolidation of these conceptualizations in Darmois' 1927 article 
\cite{Darmois27}.  Because Darmois' work was substantial and has been built on a great deal 
(this is the famous French school's contribution to mathematical relativity 
\cite{Lich44, Leray, B52, B56}... ), the timing of the appearance of Campbell's above work 
about the split is unfortunate.  Campbell's account indeed lacks this very 
useful conceptual understanding (which has subsequently become common knowledge both via the 
French school's work and the foundations of canonical GR \cite{ADM}).  

Here is the conceptualization and how it was gradually being 
discovered by Campbell's contemporaries.  At the simplest level, 
altering the ($n$ + 1)- or $n$-d signature in the split (so as to seek GR applications) is immaterial.  
[I smuggled this into the above presentation (1--3) of the split by bringing 
in an $\epsilon$ which can be set to $-1$ in the case where the $\mu$-direction is timelike; the 
hypersurface's signature does not feature explicitly in (\ref{ab}--\ref{mumu})].  There are 
two notable GR applications.  First, the split of the Lorentzian vacuum EFE's  
with respect to an entirely spatial hypersurface ($\mu = t$, a time function) is important 
because of its dynamical significance.  Second, the split of the Lorentzian vacuum 
EFE's with respect to a timelike hypersurface can be applied to 
obtaining junction conditions.  This plays no role in this paper until Sec 8.  

There are also some inter-related geometrical interpretations. 
G1: $\Omega_{ab}$ is none other than the {\it extrinsic curvature} $K_{ab}$.  
G2: the `different' equations, or {\it constraints} (\ref{amu},\ref{Gauss}), are contractions of the well-known 
{\it Gauss and Codazzi embedding conditions} \cite{GC}.  
G3: one is better served by splitting from the outset the {\sl Einstein tensor}  
${\cal G}_{AB} \equiv {\cal R}_{AB} - \frac{{\cal R}}{2}g_{AB} = 0$  and not the Ricci tensor 
${\cal R}_{AB} = 0$ whereupon the constraints (\ref{amu}, \ref{Gauss}) arise automatically as 
${\cal G}_{A\mu} = 0$.  
G4: there are `relations' between the quantities in the EFE's.
G5: These `relations' are none other than the {\sl contracted Bianchi identity}  
\be 
0 = \nabla_B{{\cal G}_A}^B 
  = \pa_{\mu}{\cal G}_A{}^{\mu} 
  + \pa_i{{\cal G}_A}{}^i 
  - {{\cal G}_A}{}^B      {{\Gamma}^{\mu}}{}_{B\mu} 
  + {{\cal G}_B}{}^{\mu}{{\Gamma}^{B}}{}_{A\mu} \mbox{ } .
\ee  
The form in G3 of the constraints makes it immediately obvious that the propagation of 
the constraints is automatically guaranteed, so there is no need to check longhand for this.  
(As regards his conjecture, Campbell's `partial proof' consisted just of establishing this 
propagation longhand, therefore he proved {\sl nothing at all}.) 
 
Finally, the inclusion of a wide range of matter types does not upset the above 
conceptualizations.  One then splits up the full EFE's
${\cal G}_{AB} = {\cal T}_{AB}$\fn{${\cal T}_{AB}$ is the energy--momentum tensor, 
whose split I denote by  
\be
{\cal T}_{AB} = 
\left(
\begin{array}{ll}
\rho & j_b \\ j_a & T_{ab} 
\end{array}
\right).
\ee
Also, I treat only simple phenomenological matter which has no further independent field 
equations.  I do this for presentational simplicity; it is well-documented that the extension 
to many sorts of more complicated matter is conceptually and 
technically straightforward.} rather than the vacuum ones.  

\mbox{ }

Hilbert in 1917 \cite{H17} was the first to recognize that the EFE's contain two systems to 
be treated separately.  He had some idea of the dynamical application and was aware of G4 
but not G5.  Levi-Civita pointed out G5 later in 1917 \cite{Levi}.  In their 1923 textbook, 
Birkhoff and Langer \cite{BL23} stated the dynamical application and explained G4--G5, 
although they did not explicitly provide the split equations.  
Also in 1923, Lanczos \cite{Lanczos23} conceived of the junction application.  
Darmois collected or discovered all of the above conceptualizations in his 1923--1927 program 
\cite{Darmois23, Darmois27}.  Hilbert and Darmois considered between them a wide range of 
matter in addition to the vacuum case.

%----------------------------------------------------------------------------------------------
\section{Early development of the GR initial value and Cauchy problems}
%----------------------------------------------------------------------------------------------

I now consider the early development of the study of the two separate systems found above 
within the EFE's: the GR IVP [solving (\ref{amu}), (\ref{Gauss})] 
and the GR CP [solving (\ref{ab})].\fn{These identifications hold in the suitable signature case 
and subject to appropriate data prescription.}  My account picks out material important for the 
discussion of Magaard's development of Campbell's conjecture in the next section.  

To prove statements like Campbell's conjecture, what one requires is what Darmois 
\cite{Darmois27} pointed out: for contact to be established between the constraint 
and evolution subsystems of the EFE's and the formal theory of p.d.e's.   
Now, the Cauchy--Kowalewskaya (CK) theorem \cite{CKlit} {\sl could} be applied here; 
after all, it can be applied to almost any type of p.d.e system.  

\mbox{ }

\noindent{\bf Cauchy--Kowalewskaya theorem}: suppose one has \sffamily A \normalfont 
unknowns $\Psi_{\mbox{\sffamily\scriptsize A\normalsize\normalfont}}$ which are functions of 
an independent dynamical variable $\mu$ and of \sffamily Z \normalfont other independent variables 
$x_{\mbox{\sffamily\scriptsize Z\normalsize\normalfont}}$.  Suppose one is prescribed on some 
domain $\Omega$ the {\sffamily A} p.d.e's of order $k$ of form 
$\frac{  \pa^{(k)}\Psi_{\mbox{\sffamily\tiny A\normalsize\normalfont}}  }
{  \pa \mu^{(k)}  } = F_{\mbox{\sffamily\scriptsize A\normalsize\normalfont}}$  where the 
$F_{\mbox{\sffamily\scriptsize A\normalsize\normalfont}}$  are functions of $\mu$, 
$x_{\mbox{\sffamily\scriptsize Z\normalsize\normalfont}}$, 
$\Psi_{\mbox{\sffamily\scriptsize A\normalsize\normalfont}}$  and of derivatives of 
$\Psi_{\mbox{\sffamily\scriptsize A\normalsize\normalfont}}$ up to 
($k$ -- 1)th order w.r.t $\mu$; furthermore suppose that in $\Omega$ the 
$F_{\mbox{\sffamily\scriptsize A\normalsize\normalfont}}$  
are \sl analytic functions \normalfont of their arguments.  Suppose one is further prescribed 
{\sl analytic data} $\Psi_{\mbox{\sffamily\scriptsize A\normalsize\normalfont}}(0, x_{\mbox{\sffamily\scriptsize Z\normalsize\normalfont}}) 
= {}^{(0)}d_{\mbox{\sffamily\scriptsize A\normalsize\normalfont}}(x_{\mbox{\sffamily\scriptsize Z\normalsize\normalfont}} )$, ... , 
$\frac{  \pa^{(k-1)}\Psi_{\mbox{\sffamily\tiny A\normalsize\normalfont}}  }
{  \pa \mu^{(k-1)}  }(0, x_{\mbox{\sffamily\scriptsize Z\normalsize\normalfont}}) = 
{ }^{(k -1)}d_{\mbox{\sffamily\scriptsize A\normalsize\normalfont}}
(x_{\mbox{\sffamily\scriptsize Z\normalsize\normalfont}})$ on some {\sl nowhere-characteristic} 
piece $W$ of the boundary of $\Omega$.  Then there \sl locally \normalfont exists a unique 
\sl analytic \normalfont solution.  

\mbox{ } 

\noindent N.B the casting of the evolution equations into CK form is robust as regards to 
what choice is made for the $n$- and ($n$ + 1)-d signature, so theorems based on the 
CK theorem would appear to have a wide range of different GR applications.   

What {\sl is} delicate however is the requirement of analyticity.  
Now, Darmois was well aware of the arguments (e.g in Hadamard's 1923 book \cite{Hadamard}) 
against the CK theorem and the use of analytic functions.  
The analytic functions are very special in the sense that theorems 
holding for these may well {\sl not} hold on other sorts of function 
spaces.  Moreover, sole consideration of the   
analytic functions is inappropriate 
for the study of `causal' or `relativistic' field equations because they are 
{\it rigid}: one cannot prescribe data independently in 
causally-disconnected regions as one would desire for physical 
modeling because the one data patch admits a unique and obligatory 
analytic continuation into the other data patch.  
Furthermore, as Hadamard points out, 
the fact that other functions can be arbitrarily-approximated by analytic 
functions is {\sl not} per se an adequate counter-argument to this.  
%
% this is not to say that analytic functions are not sometimes useful as estimates which bound functions on more general function spaces. MOVED ELSEWHERE.  
%
It would only become relevant in p.d.e problems for which the {\sl continuous dependence of 
the solution on the data can be established}.  This is in itself an important physical 
requirement, for we never know data {\sl exactly}, so we would have no physical predictivity 
unless it is guaranteed that tiny undetectable changes in the data {\sl do not immediately 
and grossly alter the behaviour of the solution.}\fn{This {\sl is} immediate from the outset, 
rather than an issue of quick onset of chaos or 
of unwanted growing modes, though well-posedness often {\sl also} bounds the growth of such 
modes \cite{RFrev}.}  
In order to have a sensible p.d.e problem in mathematical physics, it 
should be {\it well-posed}, which requires continuous dependence (and 
yet further conditions for certain types of p.d.e) in addition to 
existence and uniqueness.  Now, notably, the CK theorem is inadequate 
for establishing these important additional properties 
\cite{Hadamard, CH, Wald}.  {\sl Theorems about particular p.d.e systems 
that are based on the analytic functions alone constitute inadequate mathematical 
physics.  In the context of relativistic field equations, they also 
constitute inappropriate mathematical physics.}  
Thus such theorems should not be taken very seriously in this context.\fn{That is 
not to say that analytic functions and the CK theorem cannot play 
{\sl a part} in establishing adequate and appropriate protective theorems.  
They can, but only in conjunction with various other ingredients.  
E.g once one has a suitable norm such as in Sec 7, the analytic functions 
can in some instances be used to set a priori bounds on elements of more 
general function spaces.} 

Thus I see it as significant that while Darmois was well aware of the possibility of 
applying the CK theorem, he did {\sl not} advocate any such proof 
for the Einstein evolution equations in his 1927 article.  
While e.g Cartan did put forward a proof of this kind in 1931 
\cite{Cartan31}, note that this work does not mention analyticity nor physics but rather that 
it takes on trust that the reader is aware of the CK theorem (and presumably 
this includes the above awareness of limitations so that this work should be viewed as a purely 
mathematical exercise).  Stellmacher in 1938 \cite{Stellmacher38} was the first to successfully abandon 
the analytic functions so as to bring out the causal nature of spacetime.  The significance of 
abandoning the analytic functions was repeatedly 
stressed, e.g in Lichnerowicz's \cite{Lich44} and Four\`{e}s-Bruhat's \cite{B52} 
foundational papers on the GR IVP and CP and on p 440 of Courant and Hilbert \cite{CH}.  

It should be noted that these advances are most very specific about the 
signatures involved i.e about the $n$-d hypersurface being spacelike and 
the extra dimension being timelike.  Thus for the GR CP, inadequate 
and inappropriate analytic function results can be replaced by respectable 
theorems, but whether this is the case for any analogous problems with 
different signature is unknown (see Sec 7).  As Hilbert pointed out in 1917 
\cite{H17}, the GR CP's advances proceed via breaking manifest general covariance by 
coordinate fixings (nowadays called {\it gauge choices} \cite{YS78}) 
that are special in that the evolution equations are then cast as a 
hyperbolic system.  While the evolution equations (\ref{ab}) are 
presented above in a form close to normal gauge (which would additionally have 
$\phi$ = 1),  it is rather 
%something like 
the {\it harmonic gauge} developed by DeDonder and by Lanczos in 
1921-1923 \cite{deDonder, Lanczos22} (see also Darmois 1927) that is used to 
establish a hyperbolic system.   

\mbox{ }

The GR IVP began with Lichnerowicz 1944 \cite{Lich44}.  
He discards the normal coordinates (which are mathematically unfortunate because 
they are prone to focusing leading to their breakdown by caustic formation $K \longrightarrow 0$) 
in favour of the quite deliberately defocussing {\it maximal gauge} $K = 0$.  
He then considers the lower-dimensional metric to be known up to a conformal factor  
which is then to be determined by solving the conformally-transformed Gauss constraint.   
This started the {\it conformal method} for the GR IVP later substantially developed by York 
\cite{Yorklit, CBY80}; this 
underlies much of today's compact-binary numerical relativity \cite{numrev, Cook, BS}.  

It is furthermore relevant that Lichnerowicz's work does not yet include global considerations.  
I.e whether proposed results are local or global, and whether then they depend on the type 
of $n$-space e.g whether it is compact without boundary or asymptotically flat.  
As an example of global unawareness, it so happens that the maximal gauge cannot be 
maintained if one wishes to work globally with compact without boundary $n$-spaces; this was 
only later remedied by York's development of the {\it constant mean curvature} 
gauge ($K = C$, a spatial constant) \cite{Yorklit}.

%------------------------------------------------------------------------------------------------------------------------------------------------------------------------------
\section{Four\`{e}s-Bruhat's criticisms versus Magaard's argument}
%------------------------------------------------------------------------------------------------------------------------------------------------------------------------------

Another significant early article for the GR IVP is Four\`{e}s Bruhat 1956 \cite{B56}.  
This includes arguments for the conformal approach, favouring it over 
the following `componentwise algebraic elimination method'.  Suppose a diagonal extrinsic 
curvature component (without loss of generality $K_{11}$) is available among the quantities to 
be regarded as unknowns.  
Then one can choose to interpret the Gauss constraint as a {\sl linear} equation for this, as 
is made manifest by writing the Gauss constraint in terms of the manifestly antisymmetric 
inverse DeWitt \cite{DeWitt} supermetric $G^{ijkl} = h^{ik}h^{jl} - h^{ij}h^{kl}$:
\be
2K_{11}G^{11uw}K_{uw} + G^{1u1w}K_{1u}K_{1w} + G^{uwxy}K_{uw}K_{xy} = - \epsilon R 
\mbox{ } ,
\ee
where $u$, $w$, $x$, $y$ $\neq 1$.  Thus, provided that $Z \equiv G^{11uw}K_{uw}$ has no 
zeros in the region of interest, $K_{11}$ can be straightforwardly eliminated in the Codazzi 
constraint, which one may then attempt to treat as a p.d.e system for the other unknowns. 

Four\`{e}s-Bruhat's first criticism of the above is that this restriction 
due to the possibility of zeros of $Z$ is undesirable.  Moreover, this particular `Problem of Zeros' 
is a severe restriction because, firstly, it invalidates the 
eliminated system itself, undermining from the start any p.d.e theorems about this system.  
Secondly, since $Z$ is a function of all the metric components and of all bar $n + 1$ of the 
extrinsic curvature components, it is a function of at least one unknown so that the occurrence 
of such zeros 
within a given region is not known from the outset but rather only once the system is solved! 

Four\`{e}s-Bruhat's second criticism of the above  
is that it is a non-covariant procedure.  This contributes to it being highly ambiguous, 
since the nature of the prescription depends on the choice of coordinates, and there is no 
unique clear-cut way of choosing which components are to be regarded as the knowns and 
unknowns.  

\mbox{ }

Magaard was a differential geometry student who was not motivated by GR 
and was unaware of Bruhat's work.  He proved \cite{Mag63} Campbell's conjecture upon choosing  
analyticity and `locality' to be included in the statement: the 
{\bf Campbell--Magaard (CM) theorem}.  Of course, this is fine as regards differential geometry.  However, as argued above, the 
analyticity means that it is inappropriate to then apply this theorem as an underlying theorem for the study of 
relativistic field equations.  The analyticity is also the key reason why the theorem 
persists when one alters the signatures from Magaard's original Riemannian case to the semi-Riemannian case 
of the proposed applications.  I also argued above that 
proving existence and uniqueness alone, while useful, is inadequate 
for mathematical physics.  Magaard's use of `locality' would appear 
necessary to have a theorem at all.  Moreover, how far `local' can be extended is a complicated issue not covered by Magaard.  
But it would favour the proposed applications if a stronger meaning than the neighbourhood of analysis could be attributed to the 
word `local', as the proposed applications really concern astrophysics and cosmology and quite extended regions of spacetime 
are required to be included for these to make good sense.  
Moreover, Magaard approaches the data construction part of the proof by a particular case 
of the above componentwise algebraic method.  And yet he avoids the 
original impact of the above `Problem of Zeros' by the following sort of 
argument, although I will show later on that one then has to pay one or 
two large prices (depending on the signature) if one chooses this method and 
attempts to consider more extended regions.  

\mbox{ }

\noindent {\bf Magaard's argument}: one is entitled to declare that some $(n - 1)$-d set 
\noindent$X = \{x_1 = \mbox{ some const}\}$ is to play the role of a (partial) boundary in 
the construction of data on an adjacent $n$-d set $W$.  Then, one is entitled to prescribe 
`data for the data' on this boundary set, i.e the values that the knowns take there and a 
free choice of boundary values for the unknowns too.   Thus, although $Z$ is a function of 
knowns and unknowns in the region of interest away from the boundary set $X$, it is a 
function of knowns alone on $X$, and these knowns can be prescribed there so that $Z$ 
is bounded away from zero on $X$.   Then by continuity, there are no zeros of $Z$ 
{\sl near} $X$.  

\mbox{ }

Because Magaard's aim is to prove an embedding theorem, his particular method treats the 
lower-d metric components as knowns.  After {\bf Magaard's elimination} of $K_{11}$ by use 
of the Gauss constraint, he next treats the Codazzi constraint as a p.d.e system for the unknowns $K_{1u}$ and 
some other $K_{vw}$ component denoted $E$.  It is for each such choice of extrinsic curvature components that one has 
uniqueness.  This, and the existence, follows from the above satisfying the criteria for the CK theorem if 
one treats $K_{1u}$ and $E$ as known functions on $X$ and all the other components of 
$K_{ij}$ bar $K_{11}$ as known functions on $W$  and provided that all of the p.d.es' 
coefficients, the knowns and the `data for the data' are analytic.  So a unique solution 
exists.  

Magaard then groups \cite{Mag63, ADLR} 
%refers to my conclusion 
this data method and the usual CK local existence of a unique\fn{The uniqueness part of the proof 
is per choice of $K_{1u}$ and $E$ on $X$ and per choice of the other free components of 
$K_{ij}$ on $W$.  Clearly then there are many embeddings e.g into vacuum {\sl given the 
lower-d metric alone}.
%2 embeddings of rad universe as an example
} `evolution' to form the 
(signature-independent) CM  
theorem.  (One can proceed likewise if a cosmological constant or a large range of matter 
fields are included: the {\bf generalized Campbell--Magaard theorem} \cite{AL, DR, ADLR}).

\mbox{ }

\noindent As regards investigating whether Campbell--Magaard can hold for extended 
regions, {\bf the first price to pay} is that although the `data for the data' 
(values of $K_{1w}$ and $E$ on $X$) can be validly picked  so that $Z \neq 0$ on $X$, 
the protecting continuity argument is only guaranteed to produce a thin strip 
$W = \{0 \leq x_1 < \eta\}$ before zeros of $Z$ develop.  

And one cannot expect to be able to patch these strips together to make 
data on extended regions (`Non-patching Problem'). 
This is because the reason for terminating the construction of the 
first strip at $x_1 = \eta$ is that $Z$ is on the verge of picking up 
zeros.  So, while one can construct a second strip that extends beyond 
these zeros by choosing 
`data for the data' on 
\{$x_1 = \eta$\} that 
ensures that $Z$ is bounded away from zero there, clearly then this data 
does not match up with the first strip's data across $x_1 = \eta$.  
So what one produces is a collection of strips which generally each 
belong to {\sl different} possible global data sets.  
The evolution of each of these strips would then generally produce 
pieces of {\sl different} higher-d manifolds. 
%what I am saying is that gluing *data* is nontrivial

\mbox{ } 

It is worth making clear that the (generalized) CM theorem should not be used to claim 
that a vacuum \noindent($n$ + 1)-d manifold `surrounds' any $n$-d manifold.  All one can say is that any $n$-d  manifold can be cut up 
in an infinite number of ways (choices of the $x_1$ coordinate) into many pieces (the 
different data patches), each of which can be separately bent in an infinite number of ways 
(corresponding to the freedom in choosing both the known components of $K_{ij}$ throughout 
the relevant portion of each $W$ and the unknown components of $K_{ab}$ on each $X$), and for 
each of these bent regions one can locally find a surrounding ($n$ + 1)-d manifold e.g for 
every possible analytic function form of ${\cal T}_{AB}$.  {\sl Given that these embeddings 
are so nonunique, how can one possibly justify attaching any physical significance whatsoever 
to any particular example of embedding?}  The applications suggested in 
the literature would not seem to be very predictive \cite{ADLR} as their predictions will 
generally vary depending on the choice of embedding higher-dimensional world. 
usually give different predictions.  
N.B this sort of difficulty is by no means confined to embedding scenarios with their 
{\sl large} extra dimensions, for conceptually similar difficulties occur in compact extra 
dimension scenarios (e.g the proliferation of Calabi--Yau spaces \cite{CY} in string theory, each of which then gives 
rise to its own characteristic particle physics).  Of course, one may hope to have one day 
theoretically-motivated selection criteria to overturn this sort of difficulty.  

Finally, Four\`{e}s-Bruhat's second criticism still holds since Magaard's method lacks any 
$n$-d general covariance as it involves the choice of a coordinate $x_1$ and the 
elimination of a ${11}$-component of a tensor.

%----------------------------------------------------------------------------------------------
\section{Building more proofs: other algebraic elimination methods}
%----------------------------------------------------------------------------------------------

I first address Four\`{e}s-Bruhat's criticisms by considering two alternatives 
to Magaard's proof which I build by using distinct algebraic elimination methods.  

The `Problem of Zeros' and subsequent restrictions on region size and the 
`Non-patching Problem' are badly convoluted for the Magaard method since the zeros of $Z$ are 
not known until one 
has solved the complicated eliminated p.d.e system.  
This obstructs the further study and application of the Magaard method.  I proceed otherwise 
below.  Because the application is the proof of an embedding theorem, 
in this paper I respect the restriction to elimination methods for which the metric is a known 
(see \cite{Thanderson} for a full consideration of 3-d componentwise elimination methods).  
Consequently the Gauss constraint 
is to be viewed as an equation for some extrinsic curvature unknown, and is moreover clearly 
an algebraic equation in this.

%-----------------------------------------------------
\subsection{A second componentwise elimination proof}
%-----------------------------------------------------

First, I choose this unknown to be a nondiagonal component, which I designate $K_{12}$ without 
loss of generality.  Such a choice is always possible since the constraints require $n + 1$ 
unknowns to be properly determined, and the extrinsic curvature has only $n$ diagonal components.  
In this case, the Gauss constraint is a quadratic equation:

\noindent
\be
G^{1212}K_{12}K_{12} + G^{12cd}K_{12}K_{cd} + G^{abcd}K_{ab}K_{cd} = - \epsilon ( R + 2\rho) 
\ee
for $ab$, $cd$ $\neq 12$ or $21$.\fn{I include phenomenological matter 
into the workings of this section at no cost.}  Now, so long as 
$\bar{Z} = G^{1212} = h^{11}h^{22} - h^{12}h^{12}$ has no zeros in the region of interest, 
the solutions of the Gauss constraint may be substituted into the Codazzi constraint to form 
an eliminated p.d.e system.  Now, while there will still be a `Problem of Zeros', 
note that unlike in Magaard's method the relevant quantity $\bar{Z}$ {\sl 
is known from the start}, so one can see from the start whether the eliminated system in use
 will be valid in any particular $n$-d region that is of physical interest.  
Thus if one uses rather this method, counterexamples to the validity of the extension of 
the existence proof to cover certain regions through inapplicability of the eliminated 
system due to zeros therein may simply 
be read off: any coordinate representation of an $n$-metric for which 
$h^{11}h^{22} - h^{12}h^{12}$ has zeros will do.  

In this respect, my alternative method can be taken further than Magaard's.  
However, the Gauss equation is now a quadratic equation for $K_{12}$.  As its coefficients 
are unknown until the eliminated equations are solved, there is no guarantee that $K_{12}$ 
will turn out to be real.  So there arises a second way by which the eliminated system 
can turn out a posteriori to be invalid.  One way out of this is to use an argument like 
Magaard's.  If one starts the construction on a partial boundary set $X$, one could declare 
`data for the data' everywhere on this set such that the discriminant of the quadratic 
equation is bounded away from zero.  Then by continuity the discriminant does not become 
negative within a thin strip near $X$.

One should also point out that in both Magaard's method and my method above, additional 
conditions are required in order for the eliminated system to be castable into 
CK form.  This is an additional but yet milder version of the `Problem of 
Zeros': once the eliminated system is guaranteed to be valid in some region, there may be 
coordinate or actual conditions whereby existence and uniqueness can not be guaranteed in 
part of that region.

%--------------------------------------------------------------------------------------------
\subsection{Irreducible elimination method}
%--------------------------------------------------------------------------------------------

Second, I can do yet better by heeding Four\`{e}s-Bruhat's second criticism and abandoning 
the use of components. Like Lichnerowicz \cite{Lich44} and York \cite{Yorklit}, I work 
rather with irreducible tensor decompositions.  Unlike them, I continue to pursue an 
algebraic elimination approach to the Gauss constraint that is appropriate for embeddings 
rather than a p.d.e approach in which the $n$-metric is taken to be only partly known at the 
outset (up to a conformal factor).

In terms of the trace-tracefree decomposition  
\bea
\mbox{ } \mbox{ } 
\mbox{trace part} & K = K_{ab}h^{ab} 
\mbox{ } , \\ 
\label{t} 
\mbox{tracefree part} & 
K^{\mbox{\scriptsize T\normalsize}}_{ab} \equiv K_{ab} - \frac{K}{n}h_{ab} 
\mbox{ }, 
\label{T}
\eea
the constraints then take the form 
\bea
\mbox{ } \mbox{ } 
\mbox{Gauss} &
K^{\mbox{\scriptsize T\normalsize}}_{ij} K^{\mbox{\scriptsize T\normalsize}ij} 
- \frac{n - 1}{n}K^2 + \epsilon R + 2\rho = 0 
\mbox{ } ,
\label{Agauss} \\
\mbox{Codazzi} & 
D_b {K^{\mbox{\scriptsize T\normalsize}b}}_a - \frac{n - 1}{n}D_aK + \epsilon j_a = 0 
\mbox{ } .
\label{Acod}
\eea
I now consider the Gauss equation as an algebraic equation for the trace part $K$:
$$
K = \sqrt{\frac{n}{n - 1}}\sqrt{    K^{\mbox{\scriptsize T\normalsize}}_{ij} 
K^{\mbox{\scriptsize T\normalsize}ij} + \epsilon R  + 2\rho   } 
\mbox{ } ,
$$
and substitute this into the Codazzi equation to obtain
$$
D^iK^{\mbox{\scriptsize T\normalsize}}_{ij} - \sqrt{\frac{n - 1}{n}}D_j
\sqrt{    K^{\mbox{\scriptsize T\normalsize}}_{ij}K^{\mbox{\scriptsize T\normalsize}ij} 
+ \epsilon R + 2\rho     } = j_j 
\mbox{ } . 
$$ 
I now treat this as $n$ equations for the $n$ unknowns 
$K^{\mbox{\scriptsize TL\normalsize}}_{ab}$ encapsulated in the $n$-vector potential $W_i$:
\be
(K\delta^p_jh^{qr} - \delta^r_jK^{{\mbox{\scriptsize T\normalsize}}pq})
D_r
\left[
K^{\mbox{\scriptsize TT\normalsize}}_{pq} + 2
\left(
D_{(p}W_{q)} - \frac{1}{n}h_{pq}D_kW^k
\right)
\right]
= Kj^j - \frac{1}{2}D_j(\epsilon R + 2\rho) \mbox{ } .
\label{irredelcod}
\ee
Here $K$ and $K^{\mbox{\scriptsize T\normalsize}}_{ij}$ are treated as functions of 
$W_l$, $\pa_mW_n$ and knowns, which may be easily written down using the splits 
(9, \ref{T})
%the 10 is manual 
and  
\be
K^{\mbox{\scriptsize T\normalsize}}_{ij} = K^{\mbox{\scriptsize TT\normalsize}}_{ij} 
                                         + K^{\mbox{\scriptsize TL\normalsize}}_{ij} 
\mbox{ } ,
\ee
\bea
\mbox{ } \mbox{ }
\mbox{transverse tracefree part } \mbox{ }  K^{\mbox{\scriptsize TT\normalsize}} \mbox{ solves } 
D^iK^{\mbox{\scriptsize TT\normalsize}}_{ij} = 0 
\mbox{ } , \\ 
\mbox{longitudinal tracefree part } \mbox{ } K^{\mbox{\scriptsize TL\normalsize}}_{ij} = 2
\left(
D_{(i}W_{j)} - \frac{1}{n}D_{k}W^kh_{ij}
\right) \mbox{ } .
\eea
  
Then declaring $x_1$ to be an auxiliary independent dynamical variable, 
then if two certain functions do not have any zeros in the region of interest, 
the system (\ref{irredelcod}) may be cast into second-order CK form.  
In brief, isolating the relevant terms of the $u$-component of the Codazzi equation leads to 
\be
\pa^2_1W_u = F_u(W_i, \pa_jW_k; \mbox{ knowns}) \mbox{ } , 
\label{fstCK}
\ee
provided that division by $Kh^{11}$ is valid. 
The 1-component gives 
$$
2
\left(
Kh^{11}\frac{n - 1}{n} - K^{{\mbox{\scriptsize T\normalsize}}11}
\right)
\pa_1^2W_1 + 
\left(
Kh^{1u}\frac{n - 2}{n} - 2K^{{\mbox{\scriptsize T\normalsize}}1u}
\right)
\pa^2_1W_u
= F_1(W_i, \pa_jW_k; \mbox{ knowns})
$$
which upon use of (\ref{fstCK}) and provided that division by 
$Kh^{11}\frac{n - 1}{n} - K^{{\mbox{\scriptsize T\normalsize}}11}$ 
is valid gives 
\be
\pa_1^2W_1 = \bar{F}_1(W_i, \pa_jW_k; \mbox{ knowns}) \mbox{ } .
\ee

This provides a method of proof of Campbell's conjecture which is less prone to 
Four\`{e}s-Bruhat's two criticisms. While it has a `Problem of Zeros', it is milder than 
before in that at least the eliminated system itself is always valid, and the system is 
built in an unambiguous generally-covariant manner.

%======================================================================================================
\section{Building more proofs: topological considerations}
%======================================================================================================

The following topological considerations, left undeveloped by Magaard, are also required in 
order to understand each proof's limitations as regards the extension of its 
applicability from neighbourhoods to larger regions.  How does the 
(partial) boundary set $X =\{`x_1 = 0'\}$ in fact extend along $x_1 = 0$? What is its topology 
(open, closed, self-intersecting)?  It is also important to consider the topology of the 
$n$-d manifold, for if it is not compact without boundary, then the ultimate extension of 
the region of applicability would require treatment of boundary or asymptotic conditions.  
It may also be that the coordinate condition $x_1 = 0$ breaks down within the region of 
interest.  One can see all these issues arising from asking how far the word {\bf `local'} 
can be pushed away from its original neighbourhood meaning in order to better suit the 
proposed applications. The following issues are also similarly underpinned by what `local' 
can be taken to mean: the limitation on size of the $n$-d region on which the data is 
constructed and of the ($n$ + 1)-d evolution region.\fn{And further aspects 
in the case where the data surface is timelike, as covered in Sec 7).} 

I found that it is in fact required that the partial 
boundary set be open rather than closed or looping (self-intersecting), 
and of an analytic shape (rather than jagged or cusped).  This is 
generally so when the CK theorem is in use \cite{Hadamard}.  To 
illustrate that a closed partial boundary set is absurd, take for 
simplicity the CP for the Laplace equation.  By employing the CK 
theorem, one would then be attempting to prove that there exists a unique solution to this.  
But the problem with Dirichlet data 
$\left. \Psi\right|_{\mbox{\scriptsize X\normalsize}} = f$ {\sl already} has a unique 
solution $\Psi_{\mbox{\scriptsize D\normalsize}}$, so no solution exists for almost all 
suggestions for the rest of the Cauchy data 
$\left.\frac{\pa \Psi}{\pa \mu}\right|_{\mbox{\scriptsize X\normalsize}} = g$ 
since this is usually not equal to 
$\left.\frac{    \pa \Psi_{\mbox{\tiny D\normalsize}}    }
            {    \pa \mu    }\right|_{\mbox{\scriptsize X\normalsize}}$.   
On such grounds we require both the set for the `data for the data' and the set for the `data' to be open.  

If what one is given is precisely such a data set, then it does not 
matter exactly how far the underlying set extends. This follows from 
the allowance of a small amount of analytic continuation of the data 
beyond the edges of the set.  Thus, analytic data in a patch $X_1$ 
contains information about the analytic data in a slightly larger patch 
$X_2$ (Fig 1). Note also that the `evolution' is an analytic 
continuation and hence rigid.  So the outcome in the intersection 
of the two `evolutions' is not altered by using data patch $X_2$ 
instead of data patch $X_1$.  Note of course that there can be nearby barriers 
in some directions, so I do mean `a small amount' of analytic 
continuation in general, and the procedure whereby $X_2$ arises from 
$X_1$ cannot in general be continued ad infinitum so as to generate 
data for the whole hypersurface. 
%______________________________________________________________________
\begin{figure}[h]
\centerline{\def\epsfsize#1#2{0.4#1}\epsffile{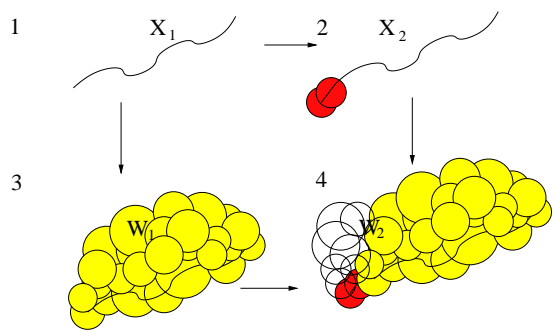}}
\caption[]{\label{TO11.ps} \footnotesize \renewcommand{\baselinestretch}{0.6}Given a set 1 on which analytic data is to be 
prescribed, if one is able to analytically extend this set by expanding as a power 
series in small balls (1 to 2), then this does not affect the `evolution' of the 
original data. (the constructed data in the yellow region $W_1$
of 3 coincides with that in the yellow region of $W_2$ in 4).\normalsize}
\end{figure}
%______________________________________________________________________
%
%*************************************JOURNAL UPDATE YELLOW TO PALE GREY ***************************************
%

Because the evolution regions are in general small, the topology of the manifolds formed by 
each evolution are not expected to enter the problem.  
%reorganize section to have less loops?***

Another complication follows from combining topological considerations with the `Non-patching 
Problem'.  Non-patching means that one cannot generally use a sequence of different 
coordinate systems that agree on their overlaps so as to construct as extensive as possible a 
region of data.  And of course, the simple case of a sphere not being coverable by a single 
coordinate system, or of the speedy breakdown of normal coordinates in general relativity due 
to caustic formation, illustrate that one generally needs to use multiple coordinate systems 
in order to cover whole manifolds.  This gives yet another reason why the theorem is about 
small patches and not whole data hypersurfaces or whole embedding manifolds.

I should also point out that, contrary to what has been hinted at in some of the 
literature \cite{singrem}, it immediately follows from the material in this section that 
Campbell's conjecture has {\sl not} been developed in such a way as to be able to say 
anything about the removal of singularities.  For, the proofs are all local whereas 
singularities are global features, and the proofs do not concern the case with boundaries 
whereas singularities are often boundaries of spacetime.  Moreover the approach to 
singularities can be rough rather than analytic \cite{Clarkebook}.  

Now the topology to be employed has been documented, pictures can be drawn of the 
three data construction methods considered in this paper and their shortcomings 
when applied to extended regions (Fig 2).
%______________________________________________________________________
\begin{figure}[h]
\centerline{\def\epsfsize#1#2{0.4#1}\epsffile{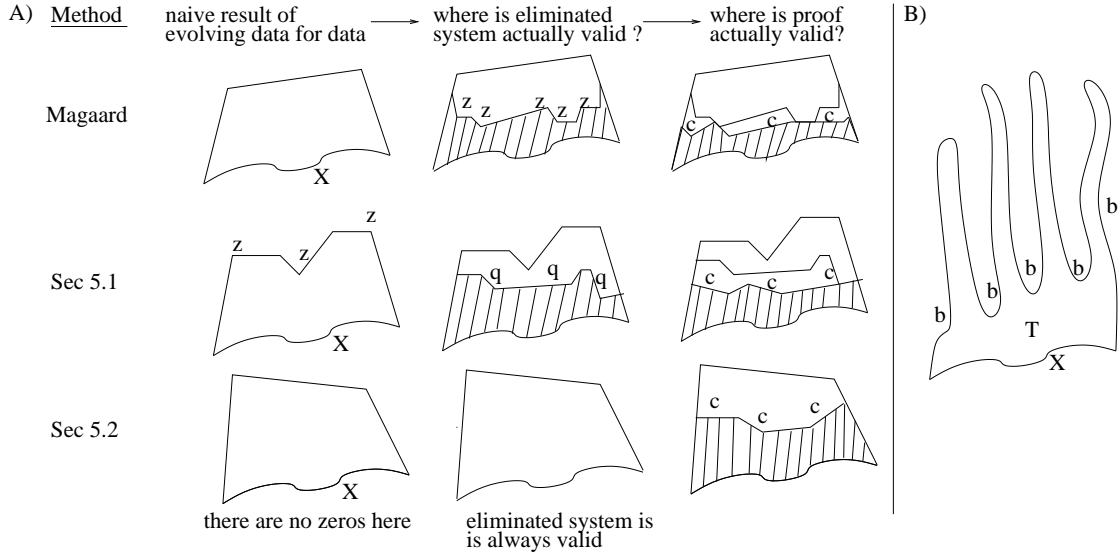}}
\caption[]{\label{TO12.ps} \footnotesize \renewcommand{\baselinestretch}{0.6} A) 
Picture of what generally 
happens in regions on which one is attempting to construct data from 
`data for the data' prescribed on a partial boundary set $X$, for each 
of the 3 data construction methods discussed in this paper.  The shaded 
region is the region $W$ for which data is being constructed by the method 
at each indicated stage.  The obstructions indicated are: zeros $z$, 
points $q$ where quadratic equations cease to have real roots and points  
$c$ where the eliminated system is not castable into 
Cauchy--Kowalewskaya form.
B) It is not yet clear at this stage how the shape of the region ought to 
be determined if one only has to contend with a bunch of {\sl isolated} 
bad points $b$ of whichever of the above kinds.  One may be tempted to 
form a more extensive `tentacled' region $T$ which nevertheless misses 
out the bad points.  I return to this issue of shape in Sec 7 for the 
timelike data surface case.  \normalsize}
\end{figure}
%______________________________________________________________________

%----------------------------------------------------------------------------------------------
\section{Difficulties if the signatures are altered}
%----------------------------------------------------------------------------------------------

%----------------------------------------------------------------------------------------------
\subsection{The domain of dependence property}
%----------------------------------------------------------------------------------------------

In the GR CP, the vast majority of serious results\fn{By this I mean results about full 
well-posedness and which are not restricted to the analytic functions (since the context is 
the general study of relativistic field equations).  It should be noted that this excludes 
all results which depend on the (otherwise very general) Cartan--Kahler theorem \cite{Bryant} 
in addition to those based on the Cauchy--Kowalewskaya theorem.} \cite{B52, Leray, B56, CB62, 
HE, HKM, CB76, FMrev, CBY80, RFrev, Klainerman} turn out to be entirely dependent 
on the signatures involved.\fn{This was first brought to attention in the `Campbell--Magaard' 
literature in \cite{ADLR}.} Not only should spacetime (of whatever dimension) have precisely 
1 time coordinate but also the data hypersurface must be spacelike for these methods to apply 
(for spacetime dimension greater than 2, see Sec 8.5). {\sl The difference between space and 
time is crucial in the p.d.e study of the EFE's.}    

For a hyperbolic-type system to be well-posed, 
in addition to existence, uniqueness and continuous dependence, 
one requires the {\it domain of dependence} property 
so that the system entails a sensible notion of causality.  
The (future) domain of dependence ${\cal D}^+[A]$ of a set of points $A$ 
is the set of points such that all 
past-inextendible causal curves through each point intersect $A$.  
Thus, if data is given on $A$, it controls the evolution in 
${\cal D}^+[A]$.  

Note that while the domain of dependence property pertains to the hyperbolic-type system itself 
and not to which boundary conditions it is equipped with, it makes a huge difference to what 
boundary conditions are sensible for the hyperbolic-type system.    
If $A$ is a piece of a spatial hypersurface, 
then ${\cal D}^+[A]$ is a reasonably extensive `wedge' or `cone' region.  
But if $A$ is a piece of a timelike hypersurface, 
${\cal D}^+[A]$ is {\sl not} a reasonably extensive region because there are usually  
causally-connected points arbitrarily close to the timelike hypersurface 
from which emanate past-directed curves which do not intersect the timelike hypersurface.  
Thus data on a timelike hypersurface for a hyperbolic-type system do not generally control 
any evolution region at all.  This suggests that conceptualizing in terms of such 
prescriptions may not make sense physically.

%-----------------------------------------------------------------------------------------------------
\subsection{Conventional mathematical physics is not available}
%-----------------------------------------------------------------------------------------------------

Still, one may have no choice as to what data one can know in a 
physical situation.  The above unfortunate\fn{One should not be alarmed by ill-posed problems 
{\sl arising}; as Tikhonov has argued \cite{Tikhonopp}, Hadamard \cite{Hadamard} was too 
extreme in suggesting no ill-posed problems would occur in nature.  Indeed, such are somewhat 
commonplace in observational sciences such as geophysics, where the observers' vantage point 
is restricted to the surface of the Earth.  Then if most of physics is postulated to be 
restricted to an apparent (3 + 1)-d hypersurface, 
it is unsurprising to run into a problem that is not known to be well-posed. The trouble is 
what to do if one is faced with such a problem!  What one should not do however is to 
{\sl trust} studies based on such problems until the mathematical physics has caught up.} situation of a 
hyperbolic-type system with data on a timelike surface is known as a {\it sideways Cauchy Problem}.  
Hitherto these did not feature very often in physical situations, giving a further impasse 
that mathematical physics for them remains greatly undeveloped.  Current results are fairly 
nonstandard, specialized and weak.  These are usually (see Sec VI.17 of \cite{CH};  
\cite{Sideways}) just for the standard wave equation, and depend on the shape of the data 
surface and often on the availability of global data for it.  While Payne's result \cite{Payne} 
is said to be extendible to 
$\Box \varsigma = F\left(x_i, t, z, \varsigma, \frac{\pa\varsigma}{\pa z}\right)$, this still 
falls short of the EFE's in harmonic gauge by not being for a system, nor for a curved wave 
operator, nor having an $x_i$-dependent $F$.  

If one believes in extra time dimensions (see e.g \cite{Hull}), one also faces a notoriously 
undeveloped problem \cite{Hewson}: the study of ultrahyperbolic equations (see Sec VI.16 
of \cite{CH} and \cite{Lavultra}).  

To appreciate this current lack of mathematical physics development, I pick out below the 
signature-specific features of proofs of well-posedness for the GR CP.  As such 
proofs are widely acknowledged as essential for guaranteeing that the GR CP rests on 
solid foundations, their absence for other signatures surely ought to be a serious concern for 
the proponents of such prescriptions in the context of higher-dimensional applications.  

In the study of hyperbolic p.d.e systems, comparison of the 
characteristic directions and the direction of integration away from 
a prescribed data surface is heavily entrenched in the works of 
Hadamard, Kirchhoff, Friedrichs, Lewy and Schauder 
%Leray is slightly different in crediting H and S and 
%improving on Petrowsky and Riesz
on which 
the works of Stellmacher \cite{Stellmacher38}, Four\`{e}s-Bruhat 
\cite{B52, B56} and Leray \cite{Leray} are based.  These first works 
relevant to GR involve in various ways the quasilinear hyperbolic form 
\be
{L}^{\mbox{\sffamily\scriptsize CD\normalsize\normalfont}}
(x,\Psi_{\mbox{\sffamily\scriptsize A\normalsize\normalfont}},
\nabla_{\mbox{\sffamily\scriptsize A\normalsize\normalfont}}
\Psi_{\mbox{\sffamily\scriptsize B\normalsize\normalfont}}) 
\nabla_{\mbox{\sffamily\scriptsize C\normalsize\normalfont}}
\nabla_{\mbox{\sffamily\scriptsize D\normalsize\normalfont}}
\Psi_{\mbox{\sffamily\scriptsize E\normalsize\normalfont}} 
= {F}_{\mbox{\sffamily\scriptsize E\normalsize\normalfont}}
(x,\Psi_{\mbox{\sffamily\scriptsize A\normalsize\normalfont}},
\nabla_{\mbox{\sffamily\scriptsize A\normalsize\normalfont}}
\Psi_{\mbox{\sffamily\scriptsize B\normalsize\normalfont}} )
\ee 
(where ${L}^{\mbox{\sffamily\scriptsize CD\normalsize\normalfont}}$ 
is a Lorentzian metric and both this and the function 
${F}_{\mbox{\sffamily\scriptsize E\normalsize\normalfont}}$ are 
differentiable a number of times and sometimes subjected to boundedness or 
Lipschitz conditions depending on exactly what one is doing).  Then e.g  
Leray's theorem \cite{Leray, Wald} holds, guaranteeing all four of 
the well-posedness criteria.  No such result is known to hold if one 
alters the signatures involved (evolving data on a spacelike surface 
with respect to a bona fide time).  

A common ingredient in these studies are energy norms. 
(These also generalize to the {\it Sobolev norms} used in the more 
modern style of proof of the four well-posedness criteria for the GR CP 
\cite{Wald, HE, HKM, CBY80, Klainerman}).     
One can see that energy norms are appropriate from simple 
consideration \cite{Wald} of e.g the flat spacetime Klein--Gordon 
equation for a scalar field $\varsigma$.  Given data on a bounded 
region $S_0$ of a spacelike hypersurface $\Sigma_0$, one can draw the 
future DOD ${\cal D}^+[S_0]$ (Fig 3a) which is the region affected 
solely by this data due to the finite propagation speed of light.  
One can then consider the values of $\varsigma$ and its first 
derivatives on $S_t = {\cal D}^+[S_0] \cap \Sigma_t, t > 0$.  Then 
using the construction in Fig 3, the divergence theorem and 
energy-momentum conservation,
\be
\int_{S_0}{\cal T}_{AC}t^{A}t^{C} + 
\int_{B}{\cal T}_{AC}l^{A}t^{C} =
\int_{S_t}{\cal T}_{AC}t^{A}t^{C} 
\mbox{ } ,
\mbox{\hspace{1.3in}}
\ee
and the second term $\geq 0$ provided that that $t^{A}$ is timelike and that 
the matter obeys the 
{\it dominant energy condition}:  
\be
-{{\cal T}^A}_Cu^C \mbox{ is a future-directed timelike or null vector } 
\mbox{ }\forall \mbox{ } \mbox{future-directed timelike } u^A.
\ee
Then the definition of the energy-momentum tensor gives
\be
\int_{S_t}\left[(\dot{\varsigma})^2 + (\pa\varsigma)^2 + m^2\varsigma^2\right] \leq
\int_{S_0}\left[(\dot{\varsigma})^2 + (\pa\varsigma)^2 + m^2\varsigma^2\right]
\label{sums}
\ee
Because each integrand is the sum of squares (each of which is 
nonnegative), this means that control over the data on $S_0$ gives 
control of the solution on $S_t$.  
If one alters the signatures so that one starts with a 
($n$ - 1)+ 1 spacetime and adds an extra spatial dimension, then there 
is no analogous notion to energy (or Sobolev) norms.  For, first there 
is generally no sideways notion of domain of dependence to make the 
construction.  Second even if we assume the higher-d dominant energy 
condition holds, it would not give an inequality because the 
perpendicular vector $z^A$ in now spacelike (Fig 3).  Third, we obtain 
a difference of squares rather than the sums in (\ref{sums}), so the 
equivalent of the energy method's use of Sobolev norms is simply of no 
use to control the `evolution' given the data.  
%______________________________________________________________________
\begin{figure}[h]
\centerline{\def\epsfsize#1#2{0.4#1}\epsffile{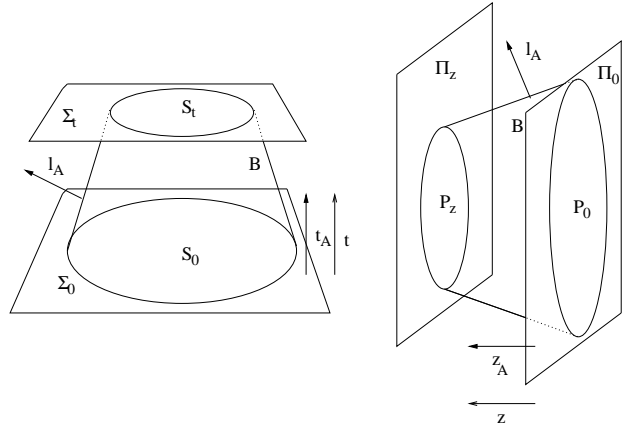}}
\caption[]{\label{TO13.ps} \footnotesize \renewcommand{\baselinestretch}{0.6} The 
bucket-shaped construction for the $n$ + 1 energy norm, 
and its generally meaningless sideways counterpart.\normalsize}
\end{figure}
%______________________________________________________________________

%-------------------------------------------------------------------------------------------------------
\subsection{The `Information leak Problem'}
%-------------------------------------------------------------------------------------------------------

{\bf The second price to pay} if one wishes to have the sideways signature 
Campbell--Magaard hold over an extended region is that 
the causal set-up is problematic.  The notion that the data on this (generally local) 
strip controls an adjacent ``evolutionary region'' is generally invalidated by an  
`Information leak Problem' (ILP), which I begin to explain in Fig 4, 
followed by further thought in the text below.  
%______________________________________________________________________
\begin{figure}[h]
\centerline{\def\epsfsize#1#2{0.4#1}\epsffile{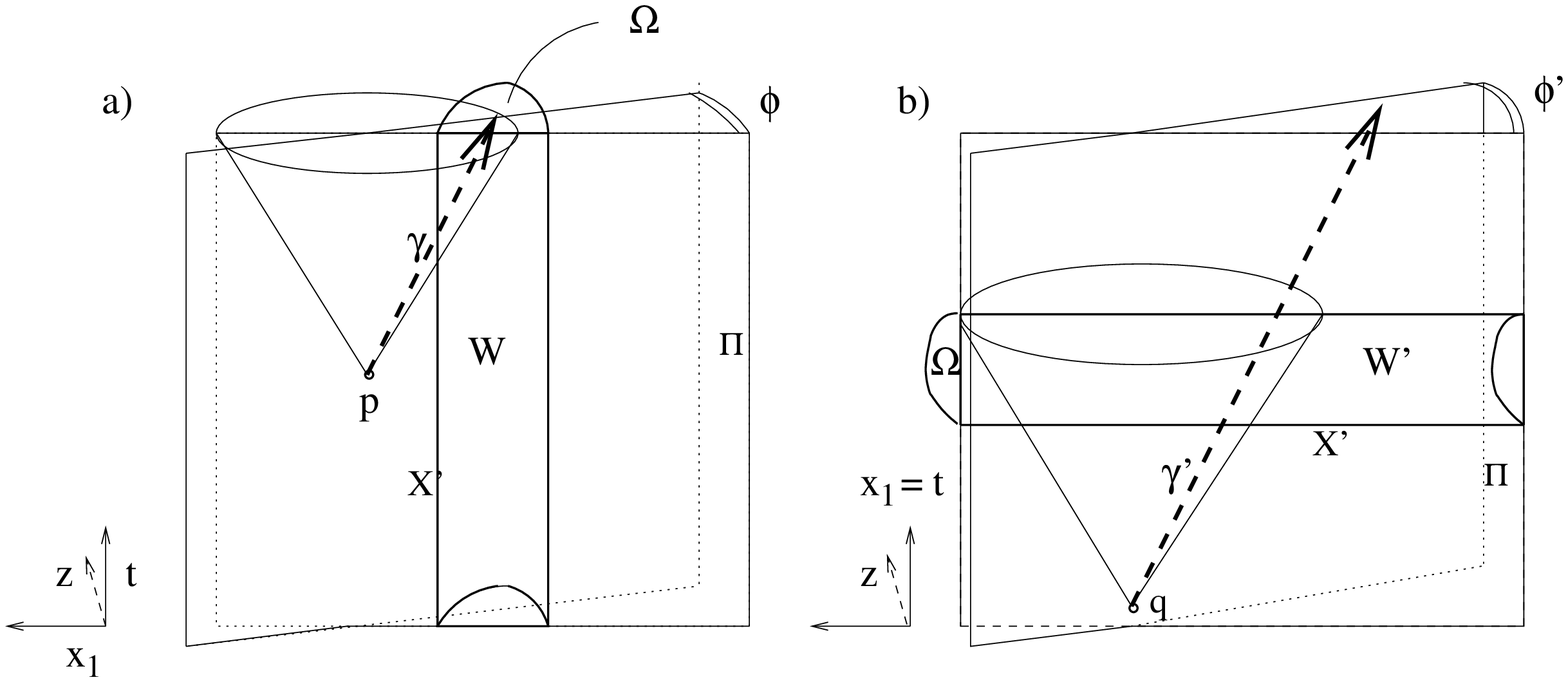}}
\caption[]{\label{TO14.ps} \footnotesize \noindent \renewcommand{\baselinestretch}{0.6}a) 
$\Pi$ is a $r$ + 1 (i.e timelike) hypersurface containing a data strip $W$ built 
by the Magaard elimination method from prescribed `data for the data' on a timelike 
$r$-d set $x_1 = 0$. The region $\Omega$ is that part of the 
($r + 1$) + 1 spacetime that is supposed to be entirely controlled by 
$W$ i.e for which Magaard's CK ``evolution" method is valid.  
Typically $W$ will not cover all of 
$\Pi$ in the $x_1$-direction, so there will be nearby points like $p$ 
outside $W$.  Consider the future null cone with apex $p$.  
No matter how thin in the $z$-direction one considers $\Omega$ to be, 
it is generally pierced by causal curves such as $\gamma$ on or in 
the null cone, by taking $\gamma$  to be at a sufficiently slender 
angle $\phi$ to $\Pi$.  Thus information can leak into $\Omega$ from 
elsewhere than $W$, which is a contradiction.  Therefore, at least in 
some places, $\Omega$ is arbitrarily thin in the $z$-direction.  One 
can envisage that sometimes there will be enough points like $p$ 
that there is no region $\Omega$ at all.  \normalsize 

\noindent \footnotesize b) Suppose $X = \{x_1 = 0\}$ is spacelike and 
is a section of the entire hypersurface $\Pi$, and $W^{\prime}$ is the 
data strip built from it by the Magaard elimination method.  There is now no room 
for points like $p$! However, typically $W^{\prime}$ will not cover all 
of $\Pi$ in the $x_1$-direction, so there will be nearby points like 
$q$ in the causal past of $W^{\prime}$ that are not in $W^{\prime}$.  
No matter how thin one considers $\Omega$ to be, it is generally 
pierced by causal curves $\gamma^{\prime}$ from $q$ \sl which do not 
pass through $W^{\prime}$\normalfont, by taking these to be at a 
sufficiently slender angle $\phi^{\prime}$.  So again, at least in 
some places, $\Omega$ is arbitrarily thin and sometimes there will be 
enough points like $q$ that there is no region $\Omega$ at all. Small 
pieces of timelike hypersurfaces need not hold useful information for 
hyperbolic-type systems! \normalsize}\end{figure}
%______________________________________________________________________

The ILP would be avoided whenever the data construction method happens 
to work globally i.e $W = \Pi$, as then there would be no points in 
$\Pi$ that are not in $W$.  But of course, general global success of 
the data construction method has not been proven! One might also 
attempt to argue that analytic continuation (the very property that 
makes the CM scheme not sensible on causal grounds) could remove the ILP.  
However, nowhere in the proofs of Campbell's conjecture has one proven that 
there exists an analytic $K_{ab}$ outside the region $W$ (and 
furthermore one whose values are compatible with those in $W$).   
Thus the data outside $W$ need not be analytic, thus analytic 
continuation is not generally applicable.  

To attain further understanding of the ILP, I note that in case a) 
the ILP {\sl already occurs on the data sheet itself} in the limit 
as the angle becomes zero.  I.e in a p.d.e problem with data in this 
configuration, the data itself suffers from non-arbitrariness of 
prescription (even if the data is {\sl not} assumed to be analytic).  
To make sense of this statement, it is helpful to consider first 
the simpler case of the sideways CP on a flat ($n$ - 1) + 1 plane for 
the $n$ + 1 flat space wave equation.  What the statement then says is 
that general strip shapes are of no use to prescribe data on, for the 
data there also imply the form the data must take in the two cones 
(drawn as wedges in Fig 5a; see p 759 of \cite{CH}).  
Additionally, one may query whether one can get rid of case b).  
Now, {\sl assuming the sole means of influence is mediation on the 
null cone}, further extension to form the double cone (depicted as a `diamond' $D$ 
in Fig 5b) successfully bans points like $q$ so that case b) would 
not occur either.  I then draw some null cones in Figs 5c and 5d to convince the reader 
that there are indeed no points 
like $p$ or $q$ {\sl as regards the sending of null signals} if one 
adopts this `diamond' construction.  Moreover, I should point out that 
the assumption that the mediation is solely on the null cone is {\sl false} if one 
has a massive wave equation or if there is a fluid flowing.  In fact it 
is even false for the plain wave equation itself for the dimension of 
interest (4 + 1) because this assumption is based on everyday (3 + 1) 
intuitions about Huygens' principle, but {\sl Huygens' principle is 
false if the number of spatial dimensions is even.}  Thus I am not 
fully overcoming the ILP, but 
rather adopting careful causal methods to understand it as best 
as possible.
%______________________________________________________________________
\begin{figure}[h]
\centerline{\def\epsfsize#1#2{0.4#1}\epsffile{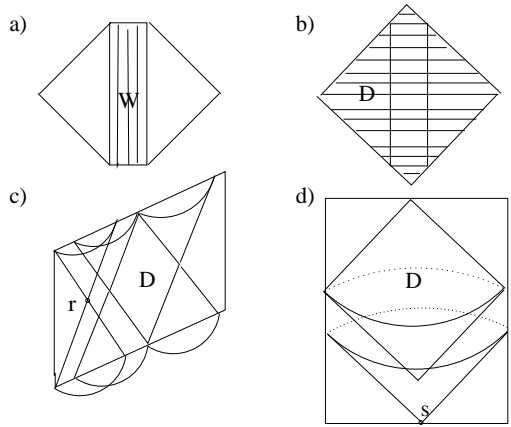}}
\caption[]{\label{TO15.ps} \footnotesize \renewcommand{\baselinestretch}{0.6} a), b) Region shapes chosen to avoid 
the `Information Leak Problem' as best as possible.  c) 
$r$ is not a point like the $p$ of Fig 4a): its cone does not interfere 
with the evolution region immediately in front of the `diamond' region $D$.
d) $s$ is not a point like $q$, at least as regards mediation on the 
null cone.\normalsize}
\end{figure}
%______________________________________________________________________

With the above insight, I now return to the more complicated case of 
the 4 + 1 EFE's and what happens to the data construction part of the 
CM theorem on a timelike hypersurface.  Starting from a set 
$X$ = \{$x_1$ = const\} on which `data for the data' is prescribed, 
I choose $x_1$ to be a bona fide time $t$ for the 3 + 1 manifold,  
and I choose to build the data by evolving the `data for the data' 
both forwards and backwards in $t$.  
This gives  preliminarily data on a `doubly thick' thin 
strip $W$ bounded on both sides by bad points (Fig 6a), whether due the 
breakdown of the eliminated system itself or of the CK proof for it.
Now, given an event (i.e a point) $m$ that the data set $W$ must 
include, build the largest `diamond' $D$ (Fig 6b) around $m$ that is 
free of bad points.  The sides of $D$ are 3 + 1 null geodesics, which 
are known since the 3 + 1 metric is a known, so at least na\"{\i}vely 
this construction can be readily made.  
I then note that, first, this construction rids us of the possibility of 
constructing data on a `tentacled' set like $T$ in Fig 2B. 
Second, that keeping only such a $D$ entails losing a great deal of the 
na\"{\i}ve strip $W$, so this construction is `even more local' than before.
%______________________________________________________________________
\begin{figure}[h]
\centerline{\def\epsfsize#1#2{0.4#1}\epsffile{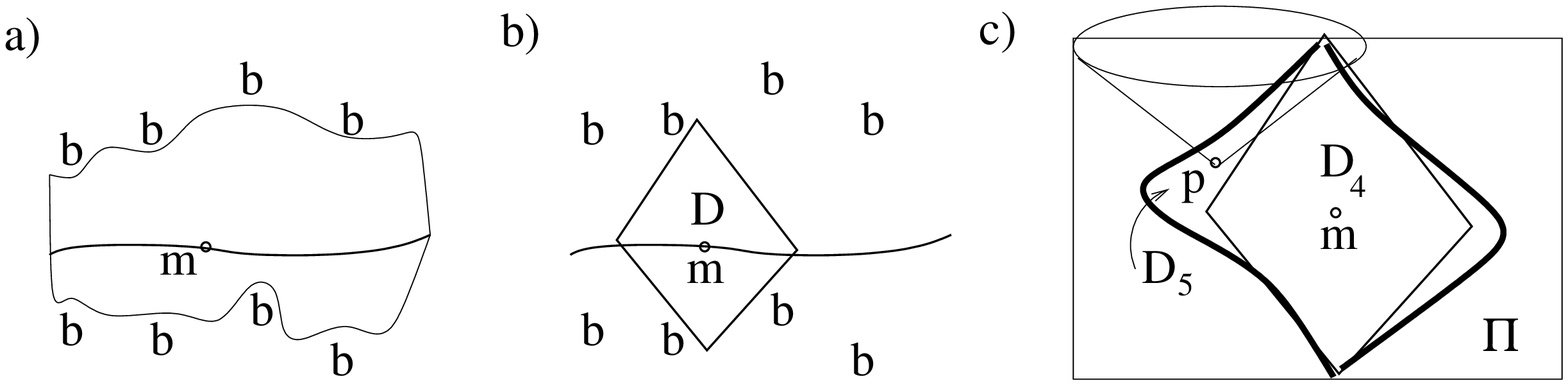}}
\caption[]{\label{TO16.ps} \footnotesize \renewcommand{\baselinestretch}{0.6} a), b) In place of a vaguely shaped strip 
$W$ I will now use a `diamond' $D$.  
c) Moreover the na\"{\i}ve 3 + 1 diamond $D_4$ may differ from the 4 + 1 diamond $D_5$, and 
points like $p$ may then reappear if one does not heed this difference.\normalsize}
\end{figure}
%______________________________________________________________________

The paragraph above about the wave equation goes to show 
that one would not expect this construction generally 
to rid us of points like $q$.  The 4 + 1 EFE's moreover give 
extra trouble not experienced in the wave equation example.  
First, 3 + 1 geodesics are not generally among the 4 + 1 ones.  
So in a straightforward 4 + 1 EFE model, the 3 + 1 null geodesics 
used to build the `diamond' are in fact generally not ascribed causal physical 
significance, while in a general braneworld model the null 3 + 1 geodesics are 
only partly relevant since causal physical significance of distinct matter types  
is ascribed both to 3 + 1 geodesics and 4 + 1 geodesics.   
In either of these situations, 
if the `diamond' $D_5$ made from 4 + 1 geodesics projected onto the 
3 + 1 world has pieces lying outside the `diamond' $D_4$ made from 
3 + 1 geodesics (Fig 6c), then points like $p$ are re-admitted.  
(The case that $D_5$ lies entirely within $D_4$ would seem to be 
less probable than $D_5$ protruding somewhere or other.)  
Furthermore, one can not a priori use $D_5$ in place of $D_4$ since 
one does not a priori know what the 4 + 1 geodesics are; their form 
only becomes apparent once the sideways evolution has been carried out 
and thus the $n + 1$ metric has been determined.  
Thus using a `diamond' construction to avoid as much of the 
`Information leak Problem' as possible for the sideways Cauchy problem 
for the 4 + 1 EFE's generally entails post-evolutionary tinkering with 
the set on which the pre-evolutionary data is prescribed.

Second, the `diamond' or `double-cone' depiction betrays na\"{\i}veness 
about what generally-relativistic null cones look like in sufficiently 
large regions.  They can be distorted by sufficiently strong gravitation 
as is well-known from black hole spacetimes.  Multiple weak focussing 
from far-apart astrophysical bodies also has an important effect on 
the cones as is well-known from the study of microlensing. This makes 
the `cones' in general \cite{EllisTavakol} infinitely complicated and 
prone to self-intersections and {\it catastrophes} such as cusp 
formation or refocussing to a point.  In Fig 7 I illustrate how each 
of these effects could render a proposed `diamond' flawed as regards its 
useability for proving CM in an extended region including a particular $m$.    
%===================================================FIGURE 7============================================================== 
%______________________________________________________________________
\begin{figure}[h]
\centerline{\def\epsfsize#1#2{0.4#1}\epsffile{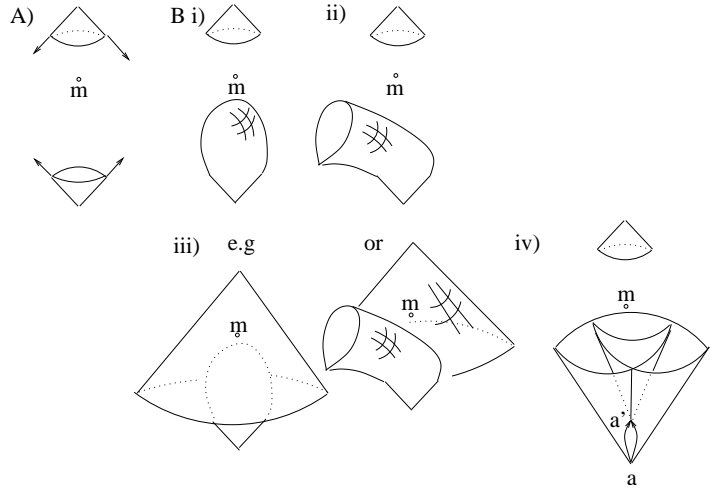}}
\caption[]{\label{TO17.ps} \footnotesize \renewcommand{\baselinestretch}{0.6} A) One is 
trying to form a `double cone' region $W$ enclosing $m$ by starting at two apices.  
B i) However, life can be cruel and one of these cones could close on itself before $m$ is surrounded by the cone.
B ii) Also one of the cones could be bent away from $m$ rather than surround $m$.
B iii) Even if these two effects do not quite prevent $m$ from being surrounded by a cone, they could totally or partially 
prevent that cone from intersecting with the other cone.  In all these cases one is failing to construct 
a $W$ that fulfils its stated purpose of enclosing $m$.
B iv) If one apex $a$ happens to have a conjugate point $a^{\prime}$ 
before the other cone is reached, then the original cone 
self-intersects.  This means that if a region $W$ successfully enclosing $m$ is formed, nevertheless it could well be 
unsuitable for our purposes though having cusps so that $W$ is no longer of use for a CK-based proof because it is 
not an analytic shape. 
B v) Fot various higher-order catastrophes the cone could suffer from, 
see \cite{Stewart}.\normalsize}
\end{figure}
%______________________________________________________________________

A way out of this complication is to resign oneself to having to 
accept increasingly small regions $W$ for data prescription.  
In the small, one could hope that the GR null `cones' 
are well-approximated by the genuine cones of SR.  
Of course however the smaller the region must be before the 
prescription makes sense, the more stringent the word `local' is in 
a theorem about that prescription.  

As argued, the ILP can in any case not generally be avoided for the 
sideways CP case of CM for an extended region.  I should stress that the ILP is an inherent 
Problem; it cannot generally be avoided by replacing Magaard's proof with one of my proofs of 
Sec 5; indeed it holds quite generally for sideways Cauchy prescriptions for the $n$ + 1 EFE's and it 
ought to be explored in this broader context with particular examples in a subsequent article.  
Moreover, of course, there is no such Problem for the bona fide CP case of CM.  Thus, 
as in the sideways case this Problem clearly affects how one ought to precisely state whichever 
variant of Campbell's theorem (see below), it is in fact not true to say that Campbell's theorem 
holds regardless of signature.  This can be encapsulated by saying that for the sideways case 
extending the word {\bf `local'} away from the neighbourhood notion 
has a series of extra connotations beyond those given in Sec 6.  Namely, to avoid as much 
of the ILP as possible by respecting causality as best as one can in a sideways problem, 
that once one has a strip one ought to form a smaller `diamond' out of it, 
that the `diamond' may require amending once the evolution has been carried out, and 
that the `diamond' may suffer additional pathologies if the region it encloses is nevertheless 
large so that it would need replacing with a smaller `diamond'.  
Also, because the ILP is nevertheless not fully avoided so that the possibility 
is opened that the embedding construction is in conflict with presupposed knowledge 
[the form of the metric outside the final data region $W = D (= D_5)$].  
One requires a new attitude to get round this, namely that one should re-declare that the 
metric was known all along only within $D$.  Note that while this enables a more carefully 
phrased version of the theorem to survive for the sideways case, as regards the applications 
of this suggested in the literature, this interpretation is merely shifting the `ILP' 
from preventing there being a theorem to tampering with its relevance to the applications 
proposed in the literature.  This is because the original intentions of the applications 
are surely the wish to embed a 3 + 1 metric well-established to be a good 
lower-dimensional model for a sizeable patch of the 3 + 1 world 
(e.g a piece of Friedmann--Robertson--Walker universe large enough for the purposes of doing 
cosmology), so one may be straying away from having proper physical motivation if one succeeds 
in embedding only a tiny portion of this 3 + 1 world {\sl and subject to pretending not to 
know what the rest of that world looks like so as to guarantee the consistency of that 
embedding}.

%=========================================================================================================================
%=========================================================================================================================
\section{Conclusion and discussion of applications}
%=========================================================================================================================
%=========================================================================================================================

%=========================================================================================================================
\subsection{The case against the Campbell--Magaard theorem}
%=========================================================================================================================

The CM theorem has been invoked as a protective theorem for the relativistic field equations of 
`noncompact Kaluza--Klein theory' (NKK) by Romero, Tavakol and Zalaletdinov 
\cite{RTZ}, especially by Wesson \cite{STMbook}, and in a number 
of subsequent articles e.g \cite{Claimsup, testsandsupp, AL, DMR, FigRom} 
(see Sec 8.3).  It and slight generalizations have also been invoked as protective theorems for 
a number of other prescriptions for the EFE's, including the construction of brane models (see Sec 8.4).  
In particular Seahra and Wesson \cite{SW03} presented the CM theorem in a heuristic manner as a protective theorem 
to underlay such constructions.   While these authors were right to seek general protective theorems, 
they have unfortunately been let down by the CM theorem being a formal differential geometry result rather than a 
result that is readily useable in relativity.    
As I have argued in this paper, from the point of view of the study of prescriptions for relativistic 
field equations\fn{Campbell's theorem might however one day be used as part of the formal study of the classification of 
GR solutions \cite{MC}.}, the CM theorem is weak because it crucially depends on the 
inappropriate analytic functions, and because the protection it provides is inadequate 
(it just concerns existence, and with additional conditions, uniqueness, rather than full well-posedness).  It is also weak to 
use the CM theorem for these applications because the interest is then really to have a theorem which holds in extended regions 
that permit meaningful astrophysical or cosmological processes to be accommodated, but the word ``local" in the statement of CM 
cannot be straightforwardly taken to cover such a notion of extended region (Sec 6--7).   This should be contrasted with the strong theorems 
in the GR CP and IVP literature (c.f \cite{B52, HKM, CB76, CBY80, Wald}): strong theorems are for {\sl all} regions, at least in many 
well-defined and substantial cases, like in the conformal IVP method, for adequate function spaces and supported by further well-posedness 
theorems.   If one seeks a theorem to serve as protection in a new application, 
and one finds a candidate theorem in the literature, it is important to check carefully whether the candidate 
theorem is adequate (strong enough to provide protection) and appropriate to the context of the application.  
Heuristic study and presentation does not suffice and can be misleading.  For example, the heuristic presentation 
in \cite{SW03} fails to convey that the CM theorem ceases to be known to be true if its statement the analytic 
functions -- inappropriate in the context the proposed application's relativistic field equations -- are replaced 
by any respectable function space.  

I also described nonuniqueness difficulties in Sec 4 and non-applicability to singular 
manifolds in Sec 6.

%==========================================================================================================================
\subsection{There is currently no replacement protective theorem for the proposed applications}
%==========================================================================================================================

In addition to arguing against the CM theorem providing protection for the suggested applications 
that involve solving relativistic field equations so as to build a higher-d world that 
surrounds our perceived 
3 + 1 spacetime from data prescribed on our apparent world, I also argue that unfortunately there are no 
known strong theorems that are appropriate for the study of this new and interesting possibility.  
To this end, I presented some general arguments in Sec 7: about how the proposed applications are causally 
absurd and amount to sideways Cauchy problems which have not been to date sufficiently studied.  
I note here that these arguments and also inconvenient locality and some forms of nonuniqueness hold against 
the completely general counterparts of the current schemes proposed \cite{311lit} for constructing bulks from 
data on branes.  

It is also important to consider whether other embedding theorems might be candidates for protective theorems.    
There are embedding theorems known for embedding into flat (Minkowski) spacetime.  
A difficulty with this is that it may appear to make little sense to trade the 3 + 1 EFE's for a 
higher-d world with no field equations at all.  Indeed, unlike in the vacuum case where CM 
was suggested to support NKK theory, there appear to be no proponents of any theories based on 
straightforward embedding into Minkowski spacetime.  
Note also that {\sl} many extra dimensions are needed to ensure general existence, potentially compounding 
nonuniqueness problems.  On the one hand, one has the Schl\"{a}fli--Janet--Cartan--Eisenhart theorem \cite{JanetCartan}, 
which has the advantage of requiring only (9 + 1)-d Minkowski spacetime to embed any 3 + 1 spacetime, 
while having the disadvantages associated with being a local theorem solely about analytic solutions.    
On the other hand, e.g Clarke's theorem \cite{Clarke} is global and requires only $C^{3}$ 
metrics, but this and all other known global embedding theorems \cite{global} involve upper bounds on 
minimum dimension required that are rather larger than those which arise in string theory.\fn{One could hope that 
these upper bounds could be lowered by technical advances.}  I should note that Katzourakis \cite{Katzourakis}, contemporarily to this article, 
has proposed a way to sew together the final regions produced by the CM theorem.  This may be of considerable interest in further developments of the topics 
in Sec 8.  

Finally, the current absence of protective theorems for the suggested applications in 
the literature means that these should be treated with caution (see however Sec 8.5 for 
some suggestions).

%==========================================================================================================================
\subsection{Concerning the foundations of `noncompact Kaluza--Klein theory'} 
%==========================================================================================================================

Next I consider NKK theory.  The field equation for this is  ${\cal R}_{AB} = 0$.
The relevance of CM would then be \cite{STMbook}
1) as a guarantee that each 3 + 1 GR solution can be represented by a 4 + 1 solution 
of ${\cal R}_{AB} = 0$.  
2) Wesson then claims that NKK theory based on ${\cal R}_{AB} = 0$ is a 
geometrization of completely general matter, so that it ``realizes Einstein's dream" of 
`transposing the basewood' of the energy-momentum tensor into the `marble' of geometry.   
  
In answer to 1), I recollect also that before CM resurfaced in the literature, Wesson already 
postulated ${\cal R}_{AB} = 0$ but admitted \cite{Wearly, W90} that it was a choice\fn{See 
also \cite{Goenner}.} among the ${\cal G}_{AB} = {\cal T}_{AB}$.  It has subsequently been 
argued and asserted \cite{RTZ, STMbook, Claimsup, testsandsupp, DMR, FigRom} 
both that CM both provides support for and specifically picks out ${\cal R}_{AB} = 0$.  
However, I have pointed out in this article that CM is not suited to provide support.  
Furthermore, CM is {\sl not a valid selection criterion} for ${\cal R}_{AB} = 0$, because there is an 
analogous statement for cosmological constants or many kinds of matter in place of the zero 
left hand side (see e.g \cite{ADLR}).  Thus a return to the earlier position that 
${\cal R}_{AB} = 0$ is a choice would seem necessary.

Note however that 2) may be interpreted as advocating further, distinct support for 
${\cal R}_{AB} = 0$.  I.e that it is the entirely geometrical character of ${\cal R}_{AB} = 0$ 
itself that is special.  But I argue below that as a prospective geometrization or 
Einstein-like unified field theory (UFT), NKK theory is not established to have several useful 
features that KK theory proper possesses.  In KK theory proper, {\sl a specific known 
fundamental matter field}, the electromagnetic potential, \sl and its corresponding field 
equations \normalfont (Maxwell's equations) are geometrized (alongside an unobserved scalar 
with its field equation).  My argument is that the slanted words above are surely a 
significant part of what a geometrization or a UFT should be about.  I.e the offering of new 
insights about known fundamental fields through representing their field equations in a novel 
geometrical way.   In contrast, NKK theory is usually considered only to include a scalar.   
Although one might \cite{STMbook}\fn{Or might not \cite{W90}, depending on interpretational 
issues.} include a more general vector field than the electromagnetic potential in addition, 
the corresponding field equation is clearly not, on inspection, that of any known vector 
field of nature.    Indeed, the 3 + 1 matter in noncompact KK theory is to be interpreted 
phenomenologically.  In all fairness, Wesson is well aware that more degrees of freedom would 
be required (see \cite{Maia, Coleycrit}) before a full physical picture could emerge from 
such a scheme.\fn{E.g Witten provided a higher-d version of KK that 
incorporates the strong and weak forces too \cite{WittenKK}.}  But it should be pointed out 
that there are two somewhat distinct further contentious issues in addition to the issue 
of counting degrees of freedom.      

\noindent A) By including fields more general than the electromagnetic potential, but not 
having enough degrees of freedom to discern an electromagnetic contribution from whatever else 
one hopes to include, one appears to be decreasing the value of the theory as a geometrization 
or UFT by losing insights specific to each fundamental field.  For example a great deal of 
quantum-mechanical success (QED, Weinberg--Salam theory, QCD) stems from quantizing 
discernible fundamental fields.  Completely switching to a NKK view of the world with its 
capacity only to include phenomenological matter would effectively amount to choosing to 
ignore these successes.  Thus it is not just a case of requiring more degrees of freedom, but 
also of requiring to supplant the phenomenological interpretation of the matter itself (or, as a long 
shot, of finding a means of recovering quantum mechanical successes from within NKK).  

\noindent B) If one attempted to argue (despite the strong caveats above) that the CM theorem 
is itself meant to have something to do with the geometrization in NKK theory, it should be 
noted that the CM theorem is a geometrical guarantee that given a 3 + 1 solution, there exists 
a unique solution of ${\cal R}_{AB} = 0$.  But such an application has a further undesirable 
feature: it is at the level of re-representing solutions piecemeal rather than at the level 
of finding and using a new geometrical representation of field equations, which would be most 
likely to be what is required to have a useful geometrization or UFT.

%==========================================================================================================================
\subsection{Comments about claimed applications in the study of embeddings and braneworlds}
%==========================================================================================================================

Despite the CM theorem being claimed to be an underlying result in a number of papers, I note that in fact  
it does not appear to be used directly in any of the constructions of specific examples contained 
within these.  Rather, the messy Magaard 
part of the prescription is simply ignored and replaced by neater methods which often\fn{It 
should be noted that there is a further method in \cite{LRRT, AL} which is 
coordinate-dependent (and not analogous to any mainstream IVP method): having $K_{ab} = 0$ 
except for a single diagonal component.} turn out 
to be among the cruder simplifications of the conformal IVP method \cite{Yorklit, CBY80, 
discoftricks}.  Unlike what is claimed e.g in \cite{SW03}, without such a specific 
prescription, all one is applying to modern scenarios is the standard embedding mathematics 
associated with the split EFE's, and not an `old but little known result'.  

Specifically, when it comes down to finding particular examples, the simplifying assumption 
$K_{ab} =  0$ has been used a lot \cite{RTZ, LRT, LRRT, Lidsey, harm, SW03}.  In the GR IVP, 
this is the {\it time-symmetric} condition and is considered bereft of much generality 
\cite{numrev, Cook, BS}.  In the current application it is rather a `space-symmetric' 
condition as regards the large extra spatial dimension, but is likewise bereft of much 
generality.\fn{It is also relevant to note that it is also the {\it totally geodesic} 
condition and the condition for absence of (brane tension) + (brane-matter energy-momentum).  
See \cite{AT3} for a discussion of these issues and references.}    

The somewhat more general assumption $K_{ab} =  Ch_{ab}$ (for $C$ constant on the 3 + 1 
hypersurface one starts on) is used in \cite{LRRT, AL, DR}.  I note that this is a subcase 
of $K = C$, which is the {\it constant mean curvature condition}, well-known in the IVP 
literature to simplify the system by decoupling (\ref{amu}) from (\ref{Gauss}) \cite{Yorklit, 
CBY80}.  A subcase of this is the {\it maximal condition} \cite{Lich44}: $K = 0$, 
which of course also generalizes the above simplifying assumption $K_{ab} = 0$, and which 
emerges in \cite{harm} as the desired condition for that application (establishing existence 
of embeddings that are also harmonic maps).  I suggest that if one is interested in more examples along the lines of the above papers, 
then a fruitful place to look is the embedding of hypersurfaces or branes of constant mean 
curvature.  For, the above decoupling technique for the constraint system carries through 
regardless of signature.  As the constraint system changes fundamentally in character in 
switching signature, specific details would have to be worked out afresh for the new 
timelike hypersurface situation.

\mbox{ } 

In the case of (thin EFE-based) braneworlds, the junction condition 
\be
K_{ab} = - \frac{1}{2}\left( Y_{ab} - \frac{Y}{3}h_{ab} \right)
\label{KY} 
\ee 
arises \cite{SMSMar}, where $Y_{ab} = S_{ab} - \lambda h_{ab}$ consists of the discontinuous 
contribution to the on-brane matter energy-momentum $S_{ab}$ and the brane tension $\lambda$.  
Now, this imposition will usually mean a conflict between what are knowns and unknowns in Magaard's method 
and what on-brane matter is of interest to the theorist.  So Magaard's method will often not be suitable.   
\cite{SW03} and \cite{DRbranes} acknowledge this but still claim support from CM while adopting 
prescriptions for the constraints other than the one for which Magaard's proof holds.    
Both these works also claim that other prescriptions for the constraints can be solved just because they can be 
declared to contain more degrees of freedom than equations.  This is non sequitur.  
It suffices to note that the system $x + y + z = 0; x + y + z = 1$ has no solutions even if $x$, $y$ 
and $z$ are all declared to be free.  Thus one needs to specifically check whether a given prescription entails 
existence.  
% this would in each case entail a caveat list like the one in this paper.  

It should also be noted that imposing (\ref{KY}) reduces the scope of what each ansatz can include.  
For example, my suggested $K = C$ ansatz implies $Y = C$.  Thus this ansatz will preclude many types of matter, 
although it will permit investigation of embeddings of interesting cases such as vacuum, phenomenological 
radiation and fundamental electromagnetism.

%=========================================================================================================================
\subsection{Some suggested routes for progress}
%=========================================================================================================================

I finish by suggesting several routes forward from this paper.  These are more modest than recent claims that the CM 
theorem is a completely general result that meaningfully underlays theories such as NKK or bulks constructions for  
braneworld scenaria.  First, as regards Magaard's method or either of my methods in this paper, 
it is after all correct to use them free of all the signature problems and proof aspects (analyticity and whatever 
kind of locality), as alternative techniques for the bona fide GR IVP.  

Second, as regards the difficulties uncovered with the general study of braneworlds, there are 
several ways forward.  One should distinguish between two distinct problems to study: 
on the one hand the bona fide dynamical computation of how a given and consistent initial 
brane-bulk configuration evolves, and on the other hand direct attempts to make predictions 
within a 4 + 1 ontology given our observations of the apparent 3 + 1 world.  
In the former case, one proceeds by constructing appropriate 4 + 0 initial data 
\cite{401N, ATpap}, and then asks what the 4 + 1 GR Cauchy problem has to say about what 
then happens to the brane(s) and to the immediately-surrounding 
bulk.\fn{`Immediately-surrounding' is meant to encompass the caveats in \cite{Cauchyhor} and in 
\cite{ATpap}.}  Such a study might also be simpler \cite{ATpap} if done for nice smooth thick branes \cite{thickbranes} rather than 
for the more habitually-studied arbitrarily-thin brane models.  
But if faced with the aspects only resolvable by the latter case, 
one may have to acknowledge that braneworlds giving rise to p.d.e problems that are not known to be well-posed.  This would mean that giving 
definite answers to what life is like on a brane would be {\sl much} harder than making comparable statements in 
ordinary 3 + 1 GR.  I do not suggest that one should jump too quickly onto the interpretation and nonstandard analysis 
of this, but rather to consider whether any optimal reformulations might arise by good use of gauge 
freedom and/or defining new variables \cite{reform}.  

Third, I note that {\sl sufficiently symmetric} problems effectively involve 
a {\sl 2-dimensional reduced bulk space}.  This special case includes as a subcase 
the embedding of homogeneous cosmologies for which the bulk metric to be a 
function of a time coordinate $t$ and a bulk coordinate $z$.  This subcase involves a reduced 1 + 1 bulk whereupon 
there is no longer a geometrical distinction between a CP and a sideways CP 
(Fig 8).  Then an analogue notion of domain of dependence, of Leray form, of 
energy integral etc, make sense in the reduced sideways problem, alongside yet 
other techniques specific to such `two independent variable' p.d.e's (see Ch 5 
of \cite{CH}).  So at least for this subcase, protective theorems are available.  
I will present this in detail elsewhere.  
Another subcase are embeddings of static spherically symmetric models for which 
the bulk metric is to be a function of a radial coordinate $r$ and a bulk 
coordinate $z$.  This subcase involves solving {\sl entirely spatial} (2 + 0) bulk 
equations so the above comments about solving spacetime equations are no longer 
applicable; on the other hand this is at the price of no dynamics being included.  
%______________________________________________________________________
\begin{figure}[h]
\centerline{\def\epsfsize#1#2{0.4#1}\epsffile{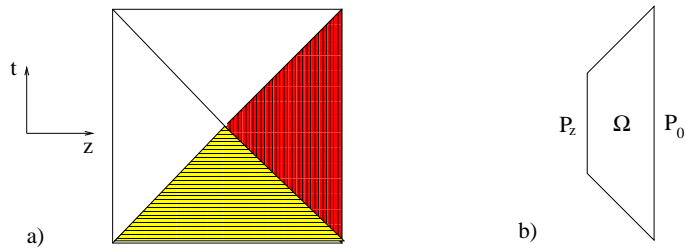}}
\caption[]{\label{TO18.ps} \footnotesize \renewcommand{\baselinestretch}{0.6} For effective 1 + 1 problems, space and time do not have a distinct formal status.  
This is because space and time are now equally complicated so one can describe $t$ as $t(x)$ just as well as one can describe 
$x$ as $x(t)$. a) The vertically striped region is now equivalent under rotation to the horizontally striped region. b) Then it 
may be interpreted that $\Omega$ is controlled by $P$ and that $\Omega$ constitutes a 
`bucket-shaped' construction which may be used in a direct analogue of energy norm proofs. \normalsize}
\end{figure}
%______________________________________________________________________

Note that while such studies would be of some use toward bulk construction prescriptions 
(indeed to date the works \cite{311lit} are restricted to static models and brane cosmology 
is restricted to at most homogeneous anisotropic models) it would be 
{\sl strictly limited to highly-symmetric models}.  Thus such work would not constitute 
anything like\fn{As examples of what one could not do within such a framework: 
1) the static case is clearly at the price of precluding dynamics.  
2) Both cases preclude a reasonably full investigation of stability and of whether the 
highly symmetric solutions are early or late time attractors.  This would include
investigating whether gravitational collapse leads to spherically symmetric black holes or investigating whether 
Friedmann--Robertson--Walker cosmology on the brane is to be expected at late times.  
3) The study of cylindrically symmetric and rotating objects, of practical importance in relativistic astrophysics, 
cannot be accommodated thus.} 
the same level of protection as was claimed from the completely general CM theorem prior to this article.  

\mbox{}

\noindent {\bf Acknowledgments}

\mbox{ } 

\noindent I thank Reza Tavakol, Carlos Romero, Julian Barbour, James Vickers, Nikolaos Katzourakis and  
especially Malcolm MacCallum for comments and discussions.  I thank Peterhouse Cambridge 
and the Killam Foundation for funding the next stage of my career.


\begin{thebibliography}{99}
\bibitem{CKlit}               The literature of the development of the Cauchy--Kowalewskaya theorem is 
                              carefully listed in the first chapter of \cite{Hadamard}.
\end{thebibliography}
\end{document}